\documentclass[journal=nalefd,manuscript=letter]{achemso}

\usepackage{chemformula} 
\usepackage{amsmath}
\usepackage{inputenc}
\usepackage{textcomp}
\usepackage{graphicx}
\usepackage{breqn}
\usepackage[T1]{fontenc} 

\author{Adarsh B Vasista}
\email{a.vasista@exeter.ac.uk}
\affiliation[Unknown University]
{Department of Physics and Astronomy, University of Exeter, United Kingdom}
\author{William L Barnes}
\email{w.l.barnes@exeter.ac.uk}
\affiliation[Unknown University]
{Department of Physics and Astronomy, University of Exeter, United Kingdom}

\title{Molecular monolayer strong coupling in dielectric soft microcavities}

\abbreviations{IR,NMR,UV}

\begin{document}

\begin{tocentry}

\includegraphics{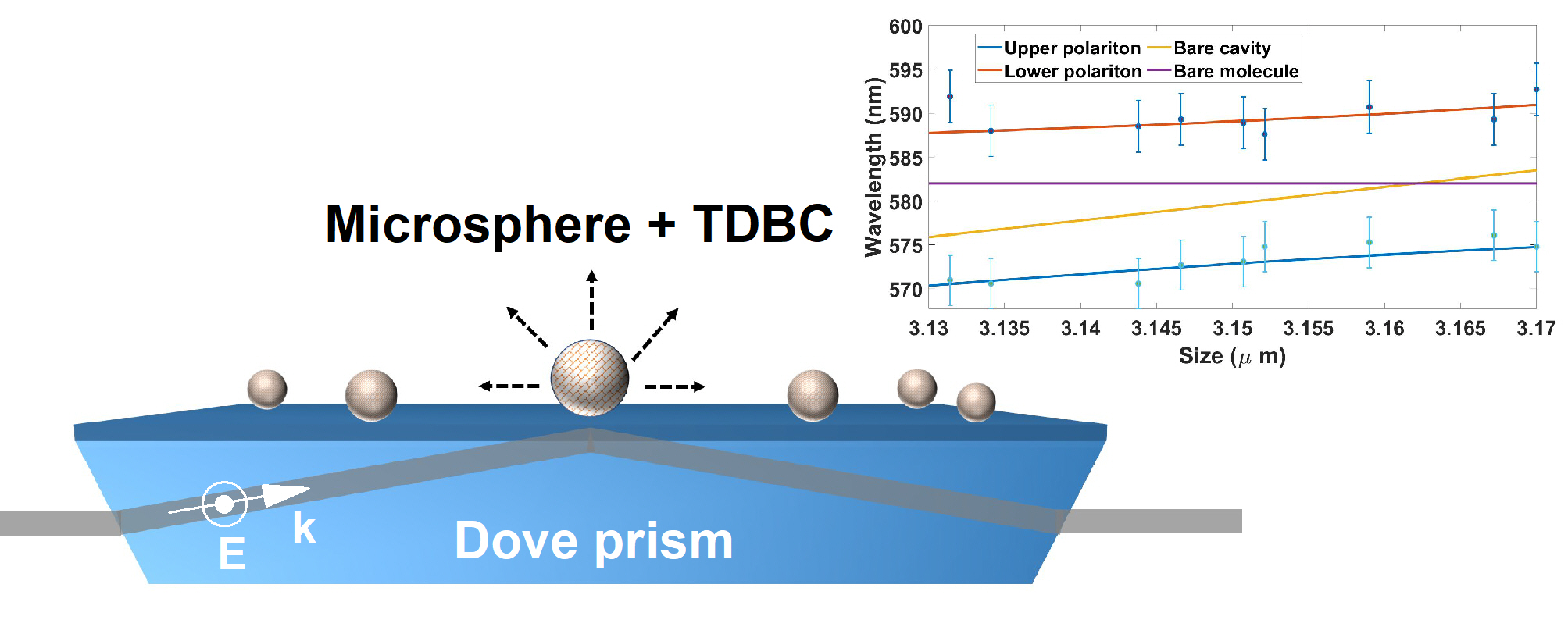}


\end{tocentry}

\begin{abstract}
  We report strong coupling of a monolayer of J - aggregated dye molecules to the whispering gallery modes of a dielectric microsphere at room temperature.
  We systematically studied the evolution of strong coupling as the number of layers of dye molecules was increased, we found the Rabi splitting to rise from 56 meV for a single layer to 94 meV for four layers of dye molecules.
  We compare our experimental results with 2D numerical simulations and a simple coupled oscillator model, finding good agreement.
  We anticipate that these results will act as a stepping stone for integrating molecule - cavity strong coupling in a microfluidic environment since microspheres can be easily trapped and manipulated in such an environment, and provide open access cavities.  
\end{abstract}

\vskip1.0cm

Strong coupling of molecules with cavity fields to create hybrid polariton states that have both molecular and photonic character has wide implications, not only in understanding fundamental concepts in physics\cite{12,13,15,16,25} but also in a plethora of applications such as polariton mediated energy transfer\cite{14,17}, selective manipulation of excited states \cite{11}, changing ground state reactivity of a molecule\cite{18}, polariton lasing\cite{27} etc.
The strong coupling regime is 
attractive because in this regime the molecular energy landscape can be radically modified, the regime thus offers great opportunities to control molecular properties\cite{Hertzog_ChemSocRev_2019_48_937,Schafer_PNAS_2019_116_4891,Ribeiro_ChemSci_2018_9_6325,Ebbesen_ACSaccounts_2016_49_2403}.
In the past, strong coupling of molecules to cavities has been explored using planar Fabry-Perot resonators \cite{20,23,14}, single plasmonic particles\cite{24,3}, meta-surfaces\cite{21,26,27}, and gap plasmonic cavities\cite{28}, among others.

Although plasmonic nanoparticles provide sub-wavelength mode volumes and significant electric field localization, making them attractive candidates to boost molecule - cavity interactions, and also belong to a class of  open cavities - where molecules can be adsorbed and desorbed easily, plasmon modes suffer from dissipative losses and typically have broad resonance spectra.
On the other hand, monolithic cavities, such as Fabry-Perot resonators, provide lower losses and narrower spectra but typically require more complex fabrication \cite{29}.
It is in this context that dielectric microspheres offer an attractive alternative.
Microspheres support spectrally sharp modes called whispering gallery modes (WGMs) that have a good degree of electric field localization\cite{30,2}.
Because of their size and composition they can be easily controlled, trapped and moved \cite{52,53}, they also provide a large surface area for molecule - cavity interactions\cite{32,33}.
Due to an ever increasing interest in microfluidics and its many applications \cite{34,35,54}, expanding the horizons of molecule - cavity coupling with mobile and controllable soft cavities, such as microsphreres and vesicles, might prove useful.

Molecular strong coupling with Mie resonances in dielectric nanoparticles have been demonstrated theoretically\cite{10,44} and experimentally \cite{9,36}.
On the other hand WGM based cavities like microdiscs and toroids have been utilized to strongly couple atoms\cite{39}, ions\cite{42}, and quantum dots\cite{40,41} at cryogenic temperatures. WGMs, in general, have been shown to exhibit exotic polarization signatures\cite{37}, orbital angular momentum\cite{46}, enhanced spontaneous emission\cite{38,45}, lasing\cite{47} etc.
Soft microcavities, like microspheres, in addition to these properties can be controlled in a precise fashion using optical forces inside a microfluidic environment.
As strong coupling of molecules to WGMs of an individual  microsphere has not been studied in detail, there is an imminent need to study strong coupling of molecules to these microcavities which will expand the horizons of its applications.
As a first step on this path we study here the coupling of molecular layers of J-aggregates to polystyrene microspheres.

In this letter we examine the evolution of strong coupling as a function of the number of deposited layers.
The signature of strong coupling was captured using single-particle dark-field scattering and the experimental results were analyzed using a coupled oscillator model, a well -established approach\cite{Wersall_ACSPhot_2019_6_2570}.
We also performed finite-element method (FEM) based numerical simulations to gain further understanding. 

\begin{figure}[h!]
\includegraphics[width=\linewidth]{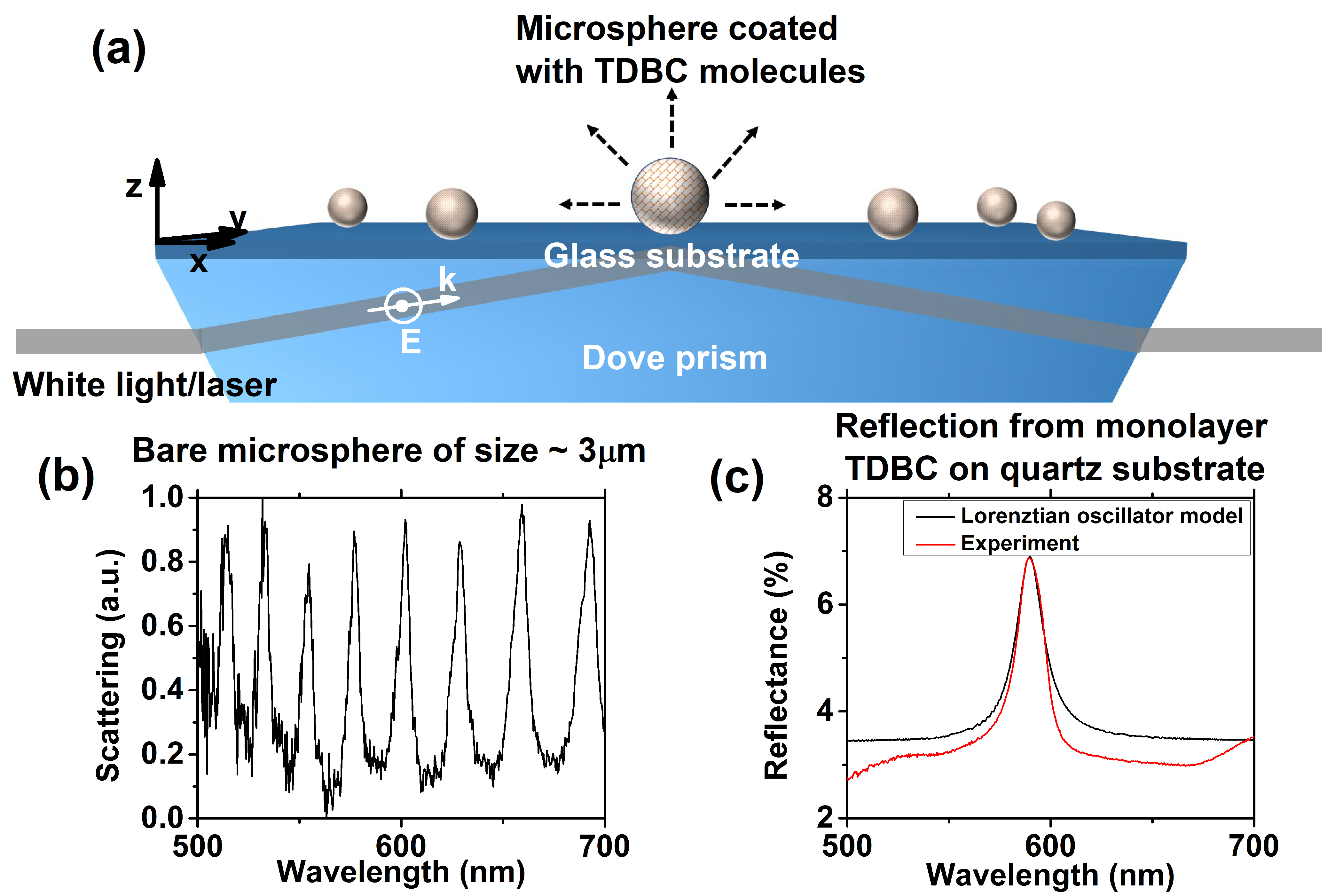}
\caption{\textit{Schematic of experiment and spectra of `bare' cavity and `bare' molecular resonance.} (a) Schematic representing the J-aggregate PDAC/TDBC coated microsphere system. Individual microspheres were probed using evanescent excitation based dark field spectroscopy. (b) Typical dark field scattering spectrum collected from a microsphere of size $\sim$3$\mu$m placed on glass substrate showing spectrally sharp whispering gallery modes. (c) Experimentally measured reflectance spectrum of a monolayer PDAC/TDBC on quartz substrate in comparison with that calculated using Lorenztian oscillator model.}
\end{figure}

Figure 1 (a) shows a schematic of the system under study. Polystyrene (PS) microspheres were coated with monomolecular layers of 5,5\textquotesingle,6,6\textquotesingle-tetrachloro-1,1\textquotesingle-diethyl-3,3\textquotesingle -di(4–sulfobutyl) -benzimidazolocarbocyanine (TDBC) J-aggregate using a layer-by-layer (LBL) deposition method\cite{1}.
In our experiments we used microspheres of two different sizes $\sim$ 3$\mu$m and $\sim$ 2.23 $\mu$m.
The LBL deposition process involved alternatively depositing oppositely charged poly electrolyte and molecular layers, further details are given in the methods section below.     

The PDAC/TDBC coated microspheres were drop-cast onto a glass coverslip and allowed to dry.
Individual microspheres were then probed using a custom-built dark-field spectroscope (see section S1 of supplementary information).
Figure 1 (b) shows a typical dark-field scattering spectrum of a bare microsphere of size $\sim$ 3$\mu$m placed on a glass substrate for TE-polarized input.
The spectral features have a linewidth of $\gamma_{cavity}$=15 meV, corresponding to $Q\sim 130$.
Figure 1 (c) shows the reflection spectrum of a monolayer of TDBC J-aggregate deposited on a flat quartz substrate.
For comparison we also show the calculated reflection spectrum for a monolayer TDBC J-aggregate using a transfer matrix method. Here the TDBC dye was modeled as a Lorentzian oscillator (see section S2 of supplementary information).
The Lorentzian oscillator model reproduces the majority of optical behavior of the LBL deposited PDAC/TDBC system\cite{51}.

\begin{figure}[h!]
\includegraphics[width=\linewidth]{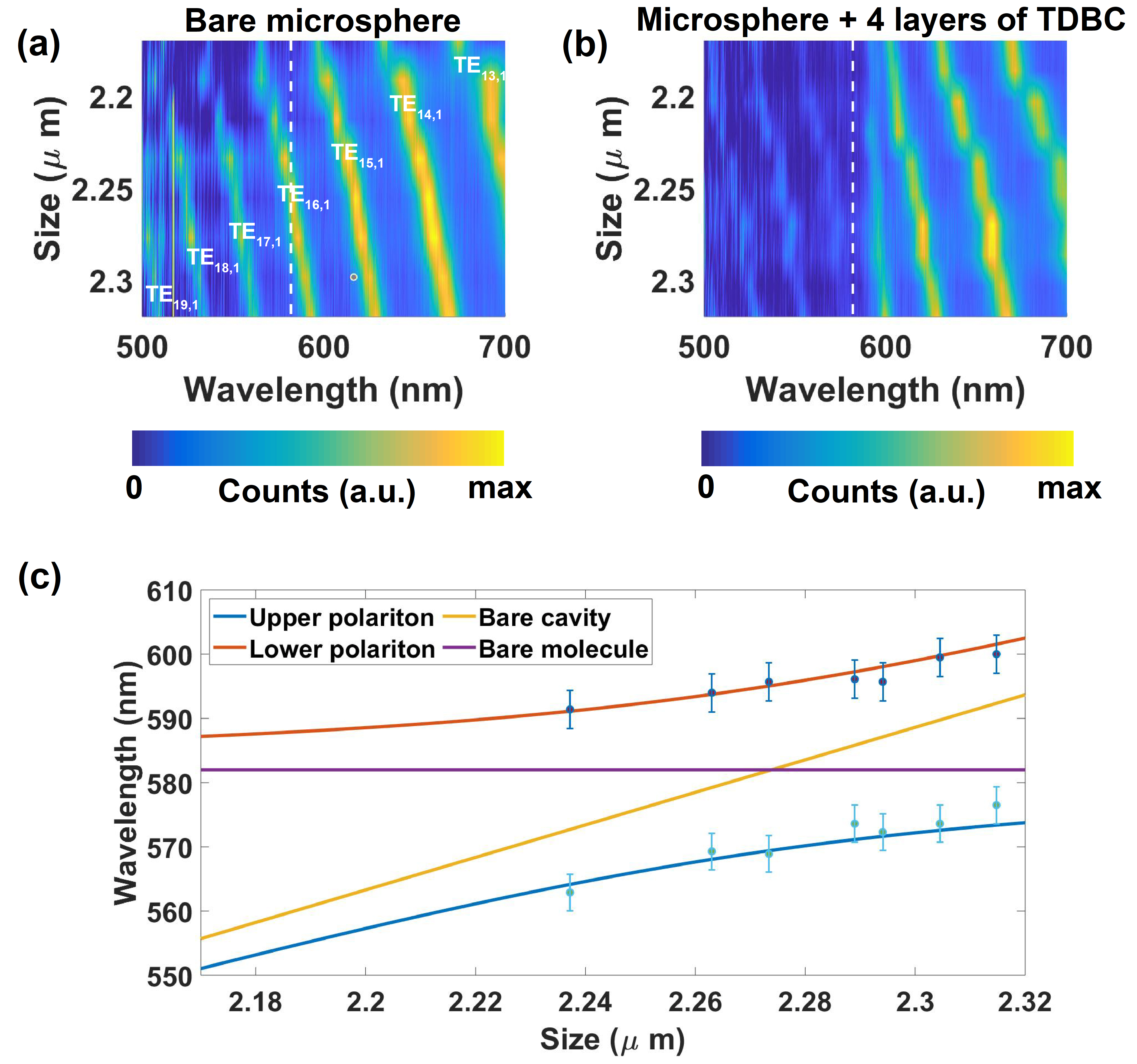}
\caption{\textit{Strong coupling of multi-layer TDBC with whispering gallery modes.}
(a) Experimentally measured dispersion of whispering gallery modes of a single microsphere of size $\sim$ 2.23 $\mu$m.
Spectra are shown as a function of microsphere size as a colour map.
The mode numbers of the WGMs are indicated on the dispersion plot.
(b) Experimentally measured dispersion of microspheres coated with four molecular layers of TDBC showing avoided crossing.
The dashed white line represents the absorption maximum of PDAC/TDBC dye.
(c) Calculated dispersion plot of the mode TE$_{16,1}$ using a coupled oscillator model to fit to the experimental data. The experimental values of both lower and upper polaritons are represented as dots.
The experimental error in determination of the polariton peak is 1\%, as indicated.}
\end{figure}

Whispering gallery resonances have strong electric field localization at the periphery of the microsphere so that the layer(s) of molecular J-aggregates coated on the microsphere experience an enhanced electric field.
WGMs supported by microspheres can be classified as transverse electric (TE) and transverse magnetic (TM) resonances.
TM modes have a radial electric field component while the TE modes have a tangential field component.
We chose to work with TE modes of the sphere and excite the microsphere with TE-polarized light (see section S3 of supplementary information). 

We began by studying multi-molecular layers rather than a single layer because the Rabi splitting ($\hbar\Omega_{R}$) for multi-molecular systems is proportional to $\sqrt N$, where $N$ is the number of molecules strongly coupled to the (cavity) field\cite{5}, the effect of thicker layers will thus be easier to see than thin layers.
We deposited four molecular layers of PDAC/TDBC on microspheres of size $\sim$ 2.23 $\mu$m. 
WGMs are morphology dependent resonances and the spectral position of the resonances are critically linked to the size of the microsphere \cite{2}. 
To understand and quantify strong coupling in multi-layer TDBC, we plotted the dispersion of WGMs for microspheres of different size, but all nominally $\sim$2.2 $\mu$m, as shown in figure 2 (a).
The sizes of each microsphere examined was determined by adjusting the size in simulated data so as to match the spectral position of the resonances, in this way we were also able to assign mode numbers as indicated.
Figure 2 (b) shows a dispersion plot of WGMs for microspheres of size $\sim$2.23 $\mu$m coated with four layers of PDAC/TDBC.
We can see the splitting and avoided crossing of the mode TE$_{16,1}$ near 582 nm.
There is a slight blue-shift in the spectral position of the avoided crossing when compared to the absorption of TDBC on a planar quartz substrate (see section S4 of supplementary information), this is due to the substrate effect \cite{4}.  

The polariton energies, i.e. the eigenenergies of the combined, hybrid system can be determined from a coupled oscillator model as follows\cite{14},

\begin{equation}
\begin{pmatrix}
E_{cavity}-i\gamma_{cavity}&-g\\-g&E_{TDBC}-i\gamma_{TDBC}
\end{pmatrix} \begin{pmatrix} \alpha \\ \beta
\end{pmatrix} = E_{pol} \begin{pmatrix} \alpha \\ \beta
\end{pmatrix}
\end{equation}

\noindent where $E_{cavity}$ is the cavity resonance energy, $\gamma_{cavity}$ is the cavity linewidth, $g$ is the strength of coupling, $E_{TDBC}$ is the molecular absorption energy, $E_{pol}$ are the polariton energies of the hybrid system, and $\gamma_{TDBC}$ is the linewidth of molecular absorption.
In the present case,$\gamma_{cavity}$=33.23 meV, $\gamma_{TDBC}$=53 meV, $E_{TDBC}$=2.13 meV.
The eigenvalues of the coupled oscillator matrix give the energies of the upper and lower polariton branches and the eigenvectors give the Hopfield coefficients\cite{50}. Figure 2 (c) shows the calculated dispersion curve of the mode TE$_{16,1}$ using the coupled oscillator model.
By fitting this model to the experimental data we determined the value of $g$ was be equal to 48 meV. The Rabi splitting, given by $\hbar \Omega=\sqrt{4g^{2}+\delta^{2}-(\gamma_{TDBC}-\gamma_{cavity})^{2}}$, was then calculated to be 94 meV.
The value of $2g$ was thus more than both the molecular and the cavity resonance linewidths, showing that we operated in the strong coupling regime.
Our system thus satisfied the criterion for the occurrence of at least one complete Rabi oscillation, which translates to $g>[\frac{1}{4}(\gamma_{cavity}+\gamma_{TDBC})= 21.55 meV]$\cite{48,49}.
The Hopfield coefficients$(|\alpha|^2,|\beta|^2)$ measure the extent of mixing of the cavity and molecular components of the polaritonic states and are shown in section S5 of the supplementary information.

\begin{figure}[h!]
\includegraphics[width=\linewidth]{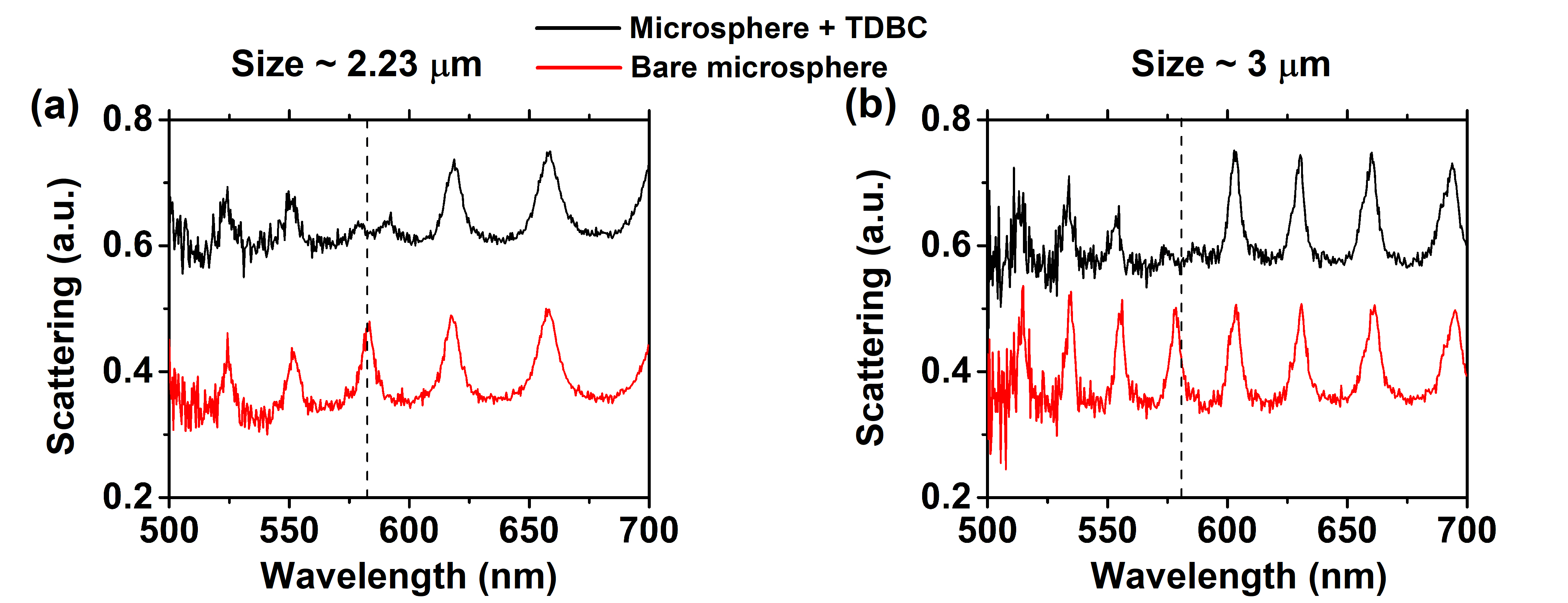}
\caption{\textit{Dark-field spectra from individual microspheres coated with a monolayer of dye.} (a) For a microsphere of size $\sim$ 2.23 $\mu$m, with and without deposition of mono-molecular layer of PDAC/TDBC. (b) Measured dark field spectra from an individual microsphere of size $\sim$ 3 $\mu$m with and without deposition of a mono-molecular layer of PDAC/TDBC. The dashed line indicates the absorption peak of the TDBC molecules.}
\end{figure}

Having established that we could see strong coupling arising from the presence of 4 molecular layers of PDAC/TDBC we next probed microspheres deposited with just a single mono-molecular layer.
Figure 3 shows representative spectra collected from microspheres of size $\sim$ 2.23 $\mu$m and $\sim$ 3 $\mu$m, both with and without the deposition of a monolayer of PDAC/TDBC.
We see a splitting of the WGM when a mono molecular layer of PDAC/TDBC has been deposited on the sphere.
The splitting of the mode is more pronounced for the $\sim$ 3 $\mu$m microsphere than for the $\sim$ 2.23 $\mu$m microsphere, something that is readily explained on the basis of larger electric field enhancement and localization, lower radiative losses provided by microspheres of size $\sim$ 3 $\mu$m.
Hence we chose to work with microspheres of size $\sim$ 3 $\mu$m to probe strong coupling of mono molecular layer of PDAC/TDBC with WGMs.
However the free spectral range of WGMs for a 3 $\mu$m sphere is quite low making this size unsuitable to study multi-layer coupling.
The evolution of the splitting of the TE$_{16,1}$ mode for different number of layers of PDAC/TDBC deposited is shown in section S6 of supplementary information.   
 
\begin{figure}[h!]
\includegraphics[width=\linewidth]{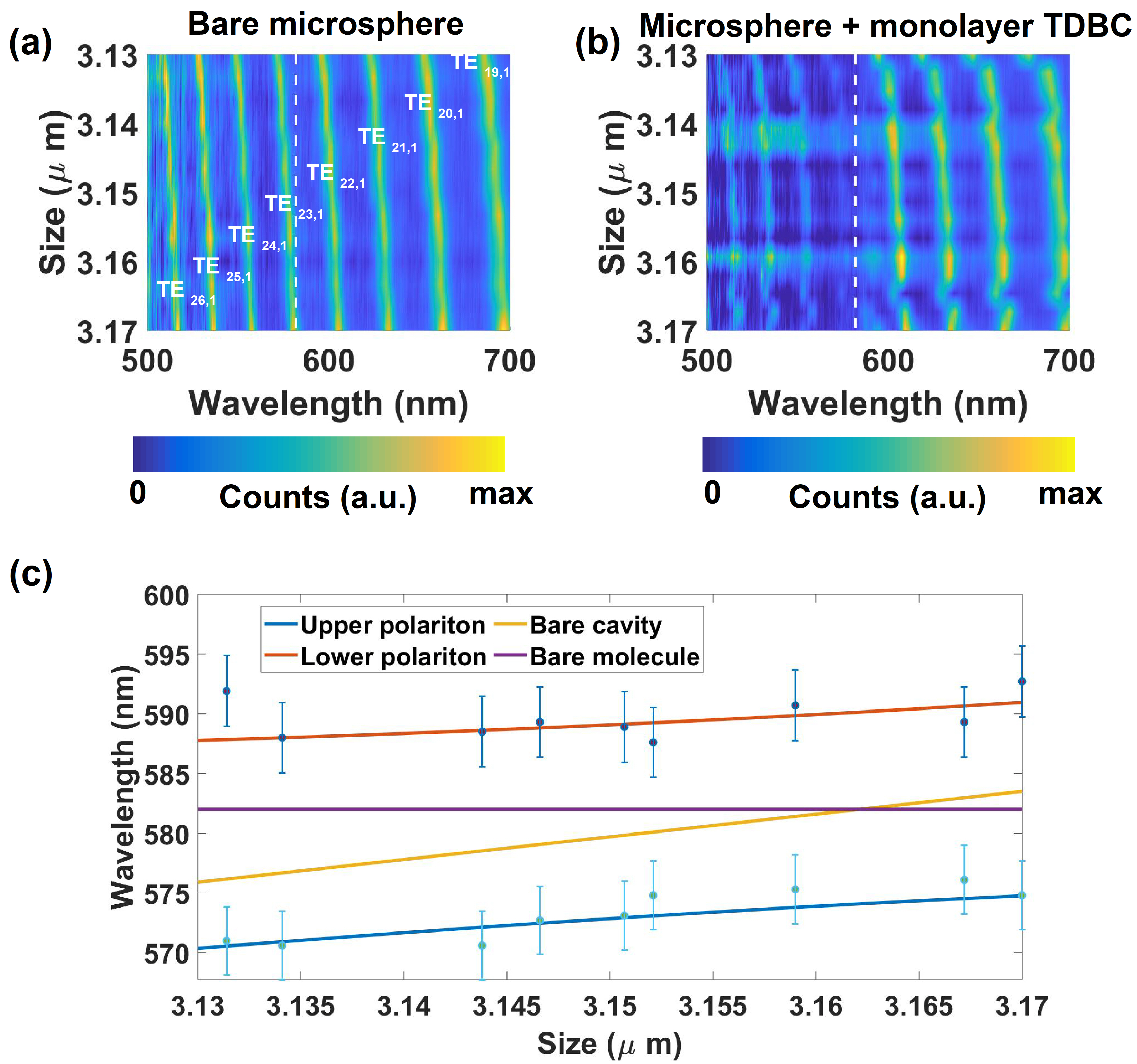}
\caption{\textit{Strong coupling of mono-molecular layer PDAC/TDBC with whispering gallery modes.}
(a) Experimentally measured dispersion of whispering gallery mode of a single microsphere of size $\sim$3$\mu$m.
(b) Experimentally measured dispersion of mono-molecular layer PDAC/TDBC coated microsphere showing avoided crossing.
The dashed white line represents the absorption maximum of TDBC dye.
The mode numbers of WGMs are indicated on the dispersion plot
(c) Calculated dispersion plot of the mode TE$_{23,1}$ using coupled oscillator model to fit the experimental data.
The experimental values of lower and upper polaritons are represented as dots.
The experimental error in determination of the polariton peak is 1\%.}
\end{figure}  

The experimentally measured dispersion of WGMs for microspheres of size $\sim$ 3 $\mu$m are shown in figure 4(a), together with the corresponding mode numbers.
The deposition of a monolayer of PDAC/TDBC on the microsphere results in a splitting of the TE$_{23,1}$ mode as shown in figure 4 (b).
To quantify the coupling we calculated the value of coupling strength, $g$, using the coupled oscillator model as described earlier.
In the case of microspheres of size $\sim$ 3 $\mu$m, $\gamma_{cavity}$ = 15 meV, and we keept $\gamma_{TDBC}$ to be as before. We can thus plot the dispersion of the mode hybridised TE$_{23,1}$ mode, shown in figure 4 (c), the value of $g$ was found to be 35 meV.
The value of the coupling strength , $2g$, was larger than both cavity and molecular linewidths, again showing that this system was in the strong coupling regime.
The calculated value of the Rabi-splitting, $\hbar\Omega_{R}$, was equal to 59 meV.
For comparison, the measured dark-field spectra for different detunings in the case of a mono-molecular layer of TDBC coated on microspheres of size $\sim$ 2.23 $\mu$m are shown in section S7 of supplementary information.

\begin{figure}[h!]
\includegraphics[width=\linewidth]{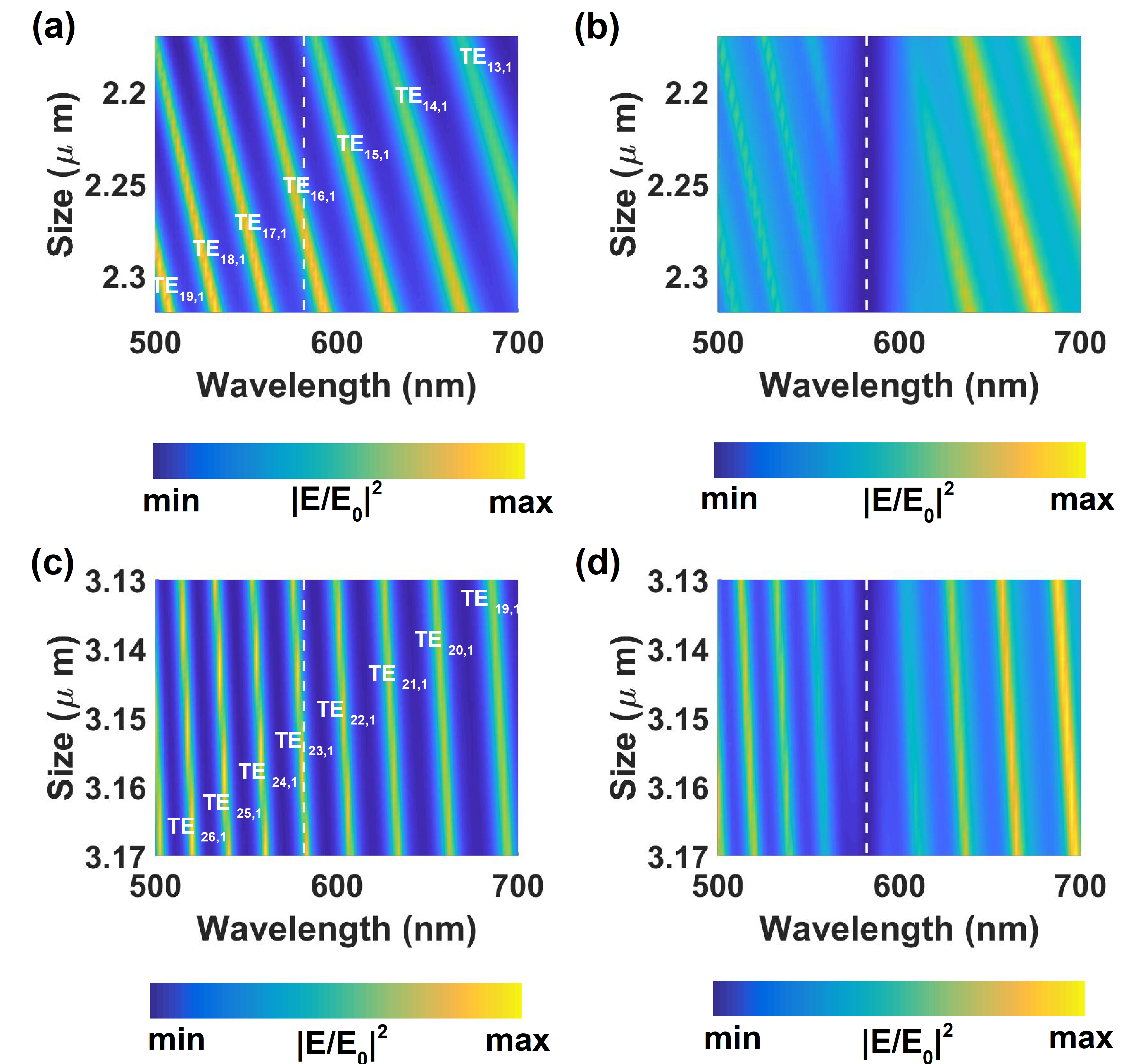}
\caption{\textit{Simulations of strong coupling with whispering gallery modes.}
(a) Numerically calculated dispersion plot of whispering gallery modes of a single bare microsphere of size $\sim$ 2.23 $\mu$m.
(b) Numerically calculated dispersion plot of microsphere coated with four molecular layers of PDAC/TDBC, showing avoided crossing.
(c) Numerically calculated plot of whispering gallery modes of a single bare microsphere of size $\sim$ 3 $\mu$m.
(d) Numerically calculated dispersion plot of microsphere coated with single molecular layer of TDBC showing avoided crossing.
The dashed white line indicates the absorption maximum of PDAC/TDBC dye.
The mode numbers of WGMs are indicated.}
\end{figure} 

To gain further understanding of the system, we performed 2D finite-element method (FEM) modeling of the hybrid system using COMSOL multiphysics\cite{55}.
The microsphere was excited using a TE-polarized broadband source in evanescent excitation configuration (see section S8 of the supplementary information for details on modeling).
Figure 5 (a) shows the dispersion of WGMs for individual microspheres of size $\sim$2.23 $\mu$m.
To model the coupling between PDAC/TDBC molecular layer and WGMs, a shell of thickness 6.8 nm (1.7 nm x 4) was introduced \cite{1}.
The resulting dispersion plot is shown in figure 5 (b). The simulated dispersion plot agrees well with the experimental data, showing an avoided crossing at 582 nm. The Rabi splitting was found to be 119 meV. The mismatch between the calculated (119 meV) and experimentally (96 meV) measured values was probably due to a combination of imperfect coverage of molecules on the microsphere, and an orientation mismatch of the molecular transitional dipole moment with respect to the polarization of the mode.
To further establish strong coupling between monolayer TDBC with WGMs we also show avoided crossing in absorption spectra of the coupled system, see section S8 of supplementary information.
We calculated the Hopfield coefficients using the coupled oscillator model with $g$=60.53 meV (corresponding to $\hbar\Omega_{R}$=119 meV), shown in section S5 of supplementary information. 

A numerically calculated dispersion plot of the WGMs supported by an individual $\sim$ 3 $\mu$m microsphere is shown in figure 5(c).
To simulate the interaction between molecular layer and WGMs we introduced a shell of thickness 1.7 nm around the microsphere (modeling details are in section S8 of supplementary information).
The resulting dispersion plot is shown in figure 5 (d), the mode TE$_{23,1}$ splits into two and undergoes an avoided crossing at around 582 nm, in agreement with the experimental results, the calculated value of Rabi splitting in this case was $\hbar\Omega_{R}$= 65 meV.
The calculated absorption spectra of the system lends further support to the system being in the strong coupling regime by showing a clear mode splitting and anti-crossing (see section S8 of supplementary information)
The Hopfield coefficients calculated using the coupled oscillator model, for the mode TE$_{23,1}$, are shown in section S5 of supplementary information.

Finally, we note that preliminary experiments to look at the photoluminescence (PL) of our systems were made.
We found that the PL spectra of our system were dependent on the mode of excitation/collection, as noted before\cite{Baieva_JCP_2013_138_044707}. Nonetheless, our data to show that some of the PL arises from the lower polariton mode, details are presented in section S9 of supplementary information.  

In summary, we have shown strong coupling between a single dye monolayer and the whispering gallery modes of a soft dielectric cavity.
We studied the evolution of the strong coupling as the number of mono-layers of our dye system, PDAC/TDBC, was increased, finding that the splitting increased, as expected.
We supported our experimental results with numerically simulated data.
This study provides a new launch point for polaritonic chemistry by providing a new open cavity system that is ideally suited to microfluidic systems and applications.
Furthermore, the microsphere dielectric cavities we have employed should be well suited for optical trapping and thus manipulation.
The results presented here may also be extended to study strong coupling between transition metal dichalcogenides (TMDs) and microspheres supporting modes with angular momentum.

\section{Methods}
For the layer-by-layer fabrication we used a cationic poly(diallyldimethylammonium chloride) (PDAC) solution as the polyelectrolyte binder for anionic TDBC J - aggregate solution.
A typical deposition step consists of mixing 20 $\mu$l PDAC solution (20$\%$ by weight in water - diluted 1:1000) with 1 ml of PS microsphere colloidal solution (15$\%$ by weight in water - diluted by 50 times) and allowing to settle for 20 minutes.
The solution was then washed 3 times in water to remove excess polyelectrolyte solution.
This step forms a positively charged layer of PDAC on microspheres.
Then 100 $\mu$l TDBC solution (0.01 M) was added to the microsphere/PDAC colloid and kept for 20 minutes.
The solution was then washed 3 times in water to remove excess TDBC molecules.
This procedure was repeated to deposit multiple layers of J-aggregate TDBC on microsphere surface.
Finally the TDBC layer was protected by depositing a layer of PDAC molecules.

\begin{acknowledgement}

ABV would like to thank Wai Jue Tan for his help in preparing samples. The authors acknowledge the support of European Research Council through the Photmat project 
(ERC-2016-AdG-742222
:www.photmat.eu). ABV acknowledges fruitful discussions with Rohit Chikkaraddy during the initial stages of the work. 

\end{acknowledgement}

\begin{suppinfo}

A document containing information about simulation strategy and dark field scattering spectra from microspheres are available free of charge.

\end{suppinfo}

\bibliography{ref_nl}

\end{document}